\documentclass[aps,prl,showpacs,twocolumn,amsmath,amssymb]{revtex4}
\usepackage{graphicx}
\usepackage{Jonasmacros}
\usepackage{bm}
\begin{document}
\date{\today}
\title{Dynamical exchange interaction between localized spins out of equilibrium}
\author{J. Fransson}
\email{Jonas.Fransson@fysik.uu.se}
\affiliation{Department of Physics and Astronomy, Uppsala University, Box 530, SE-751 21\ \ Uppsala}

\begin{abstract}
The electron mediated exchange interaction between local spins adsorbed on two-dimensional surface is studied under non-equilibrium conditions. The effective spin-spin interaction is found to depend both on the spin-polarization of the substrate and the excitation spectrum of the local spins. For spatially anisotropic spin-polarization of the substrate, the spatial dependence of the interaction comprise components decaying as $\sin(2k_FR)/(2k_FR)$ and $\sin(2k_FR)/(2k_FR)^2$.
\end{abstract}
\maketitle

The excitation spectra of spin systems strongly depends of the type of interactions that are involved. The magnetic moment of e.g. single Co \cite{gambardella2003,wahl2007,balashov2006,meier2008,otte2008,balashov2009,otte2009,zhou2010}, Fe \cite{hirjibehedin2007,yayon2007}, Cr \cite{yayon2007}, and Mn \cite{hirjibehedin2006} atoms become strongly anisotropic due to symmetry reduction in the interaction with electron medium. Studies of inelastic scattering processes of layered materials \cite{balashov2006} and single atoms \cite{meier2008,otte2008,balashov2009,otte2009,zhou2010,hirjibehedin2007,yayon2007,hirjibehedin2006,fransson2009,loth2010} have given deepened insight to the excitation spectra of various elements, which then provide further detail to the understanding of the involved interactions.
phys
For magnetic systems, the interactions between the local spins can be of different character, which is often modeled using e.g. the Ising and Heisenberg Hamiltonians, but also anisotropic models such as e.g. $XY$- or anisotropic Heisenberg Hamiltonians. Regardless of model, the interaction parameters describe a physical interaction between the spins, which result from different mechanisms. The spin-spin interaction may be direct in the sense that the exchange Coulomb integral ($\int\psi^\dagger_\sigma(\bfr)\psi^\dagger_{\sigma'}(\bfr')V(\bfr,\bfr')\psi_\sigma(\bfr')\psi_{\sigma'}(\bfr)d\bfr d\bfr'$) is non-negligible, or of indirect nature, e.g. super-exchange or double-exchange. Of particular interest is the Ruderman-Kittel-Kasuya-Yosida (RKKY) interaction \cite{ruderman1954,kasuya1956,yosida1957}, which is generated by a coupling between the local spins and the surrounding electron medium, such that the spin-spin exchange interaction is mediated by the electronic environment.

In this paper, we address the electron mediated exchange interaction between localized spin under non-equilibrium conditions in two-dimensional systems. The question is pertinent to recent measurements using e.g. scanning tunneling microscopy (STM) where local non-equilibrium conditions are created by the tunneling current. It is demonstrated that the resulting spin-spin exchange interaction depends on the spin-polarization of the electron medium and on the excitation spectrum of the localized spins. For spatially anisotropic spin-polarized surface electrons the spin-spin interaction comprise the Ising, Heisenberg, and Dzyaloshinski-Moriya interactions, where the latter is shown to asymptotically decay as $\sin(2k_FR)/(2k_FR)$. In contrast to previous studies of the non-equilibrium RKKY interaction \cite{schwabe1996,chen2010}, we here also include the proper time-dependence of the local spins.

The presence of the local spin (or magnetic) moments, results in a spatially inhomogenous surface electron spin-polarization which can be transformed into a spatially non-uniform spin bias distribution between the spin-projections of the surface electrons. Under the spin biased conditions, electrons flow between the local spin moments, however, different spin projections travel in different directions, thus, establishing a net equilibrium. The setup is, hence, reminiscent of the electron spin resonance situation discussed in Refs. \onlinecite{zhang2003,fransson_zhu2008}, and is ideal for investigating the electron mediated exchange interaction in terms of non-equilibrium formalism.

We begin by considering localized spins $\bfS_\bfr$ at the positions $\bfr$ interacting with a continuum, treated in the closed time-path Green function formalism \cite{chou1985} which was recently applied to spin dynamics in a Josephson junction \cite{zhu2004} and three-dimensional metallic systems \cite{onoda2006}. We calculate the partition function (in units: $\hbar=c=1$)
\begin{subequations}
\begin{align}
{\cal Z}[\bfS_n(t)]=&\tr T_Ce^{i\calS},
\\
{\cal S}=&{\cal S}_\text{\tiny{WZWN}}
	+{\cal S}_\text{ext}
	+\oint_C[\Hamil_K+\Hamil_T]dt,
\\
\oint_C(\cdot)dt=&\int_{-\infty}^\infty(\cdot)dt_+-\int^\infty_{-\infty}(\cdot)dt_-,
\end{align}
\end{subequations}
where we have omitted unimportant contributions from the electron gas with quadratic dispersion and isotropic effective mass $m$, as we are considering conduction electrons in the continuum approximation. The STM tip is assumed to have negligible effect on the spin cluster. ${\cal S}_\text{\tiny{WZWN}}=\sum_r\int\bfS_r(t)\cdot[\bfS_r(t)\times\dot{\bfS}_r(t)]dt/S^2_\bfr$, $S_\bfr=|\bfS_\bfr|$, is the Wess-Zumino-Witten-Novikov (WZWN) term describing the Berry phase accumulated by the local spins. The trace runs over the degrees of freedom for the electrons in the tip and substrate in order to provide an effective spin action, which in the present situation represents the interaction of the magnetic spins with a non-equilibrium environment. $\calS_\ext$ represents the coupling between the system with the external electromagnetic field. The Hamiltonians inside the contour integral define the (Kondo) coupling between the local spins and the surface electrons, $\Hamil_K=-v_uJ_K\sum_\bfr\bfS_\bfr(t)\cdot\bfs(\bfr,t)$, and the coupling to external electrodes $\Hamil_T$ which generates a tunneling current in and/or out from the two-dimensional surface. For example, recent STM measurements motivates to model the tunneling current between the tip and surface \cite{fransson2009} using $\Hamil_T=\sum_\bfr\sum_{\bfp\bfk\sigma\sigma'}\cdagger{\bfp}(\delta_{\sigma\sigma'}T_0+T_1\bfsigma_{\sigma\sigma'}\cdot\bfS_\bfr)\cs{\bfk\sigma'}e^{i\bfk\cdot\bfr+i\phi(t)}+H.c.$, where  $\bfp\ (\bfk)$ denotes the momentum for electrons in the tip (substrate), whereas $T_0$, $T_1$ are the ($\bfp$ and $\bfk$ dependent) rates for the direct and exchange coupled tunneling. Here, $\bfs(\bfr,t)$ is the electron spin density, whereas $v_u$ and $J_K$ defines a unit surface element and the Kondo coupling to the electrons. $\phi(t)=e\int_{-\infty}^tV_{sd}(t')dt'$ gives the energy shift due to the bias voltage $V_{sd}(t)$ applied between the tip and surface and $\bfsigma$ is the vector of Pauli spin matrices.

The procedure in \cite{zhu2004,onoda2006,franssonNJP2008} yields the effective action
\begin{align}
{\cal S}=&
	{\cal S}_\text{\tiny{WZWN}}
	+\int\sum_\bfr[g\mu_B\bfB(\bfr,t)+\bfj^{(1)}(\bfr,t)/e]\cdot\bfS_\bfr^2(t)dt
\nonumber\\&
	-(v_uJ_K)^2\int\sum_{\bfr\bfr'}\bfS_\bfr^2(t)\cdot{\cal F}^r(\bfr,\bfr';t,t')\bfS_{\bfr'}^1(t')dtdt'
\nonumber\\&
	+\frac{1}{e}\int\sum_{\bfr\bfr'}\bfS_\bfr^2(t)\cdot\bfj^{(2)}(\bfr,\bfr';t,t')\bfS_{\bfr'}^1(t')dtdt',
\label{eq-Saction}
\end{align}
where $\bfS^1(t)=[\bfS(t_+)+\bfS(t_-)]/2$ and $\bfS^2(t)=\bfS(t_+)-\bfS(t_-)$, whereas $\calF^r_{ij}(\bfr,\bfr';t,t')=(-i)\theta(t-t')\*\av{\com{s_i(\bfr,t)}{s_j(\bfr',t')}}$ is the retarded spin GF of the surface electrons. $\bfj^{(1)}(\bfr,t)=j^{(1)}(\bfr,t)\hat{\bf z}$ and $\bfj^{(2)}(\bfr,\bfr';t,t')$ are the spin-polarized current density and spin current density, respectively, between the tip and the substrate generated by the spin-imbalance and non-equilibrium conditions in the electrodes \cite{franssonNJP2008,fransson2008}. This, general, formulation of the action is motivated from the perspective of recent tunneling experiments. In this paper the focus, however, is on the third term to the right in Eq. (\ref{eq-Saction}), which represents the RKKY interactions as it emerges from the (Kondo) coupling between the localized spin moments and the surface electrons.

Owing to the general non-equilibrium conditions, the retarded spin GF is expressed in terms of the lesser and greater surface electron GFs $\bfG^{\stackrel{\scriptstyle <}{>}}(\bfr,\bfr';t,t')$, that is,
\begin{align}
\calF^r_{ij}(\bfr,\bfr';t,t')=&
	(-i)\theta(t-t')\tr_S
	[
		\sigma^i\bfG^>(\bfr,\bfr';t,t')
\nonumber\\&\times
		\sigma^j\bfG^<(\bfr',\bfr;t',t)
		-\sigma^i\bfG^<(\bfr,\bfr';t,t')
\nonumber\\&\times
		\sigma^j\bfG^>(\bfr',\bfr;t',t)
	],
\end{align}
where the trace $\tr_S$ is taken over spin space of the surface electrons. For non-interacting but spin-polarized surface electrons we write the real space GFs according to $\bfG^{\stackrel{\scriptstyle <}{>}}(\bfR;\tau)=\int\bfG^{\stackrel{\scriptstyle <}{>}}(\bfk;\tau)e^{i\bfk\cdot\bfR}d\bfk/(2\pi)^2$, where $\bfR=\bfr-\bfr'$ and $\tau=t-t'$. The lesser and greater forms of the GF can be written
\begin{align}
\bfG^{\stackrel{\scriptstyle <}{>}}(\bfk;\tau)=&
	\sum_\sigma(\sigma^0+\bfsigma\cdot\bfDelta\sigma^z_{\sigma\sigma})G_\sigma^{\stackrel{\scriptstyle <}{>}}(\bfk;\tau)/2,
\label{eq-Gfunc}
\end{align}
where $G_\sigma^{\stackrel{\scriptstyle <}{>}}(\bfk;\tau)=(\pm i)f(\pm\leade{\bfk})\exp(-i\leade{\bfk}\tau)$ with $\leade{\bfk}=\dote{\bfk}+\sigma|\bfDelta|/2$, whereas $\sigma^0$ is the identity matrix, and $f(x)$ is the Fermi function. The effective spin-splitting $\bfDelta$ is generated by the surface electrons due to coupling to internal and/or external spin degrees of freedom. $\bfDelta$ can partially be due to e.g. the spatially inhomogeneous mean field $-v_uJ_K\sum_\bfr\av{\bfS_\bfr}$ which is generated by the adsorbed spins, and partially due to e.g. spin-orbit interactions in the surface, pertinent to recent STM measurements of Co/Pt(111) \cite{meier2008}. Using Eq. (\ref{eq-Gfunc}) and the identity $(\bfA\cdot\bfsigma)(\bfB\cdot\bfsigma)=(\bfA\cdot\bfB)\sigma^0+i(\bfA\times\bfB)\cdot\bfsigma$ \cite{sakurai} we find, after some algebra,
\begin{align}
\bfS_\bfr^2&(t)\cdot\calF^r(\bfR;t,t')\bfS_{\bfr'}^1(t')=
\nonumber\\=&
		\frac{1}{2}\sum_{\sigma\sigma'}
		F_{\sigma\sigma'}(\bfR;\tau)
		\Bigl\{
			2\sigma^z_{\sigma\sigma}\sigma^z_{\sigma'\sigma'}
			[\bfS_\bfr^2(t)\cdot\bfDelta][\bfS_{\bfr'}^1(t')\cdot\bfDelta]
\nonumber\\&
			+[1-|\bfDelta|^2\sigma^z_{\sigma\sigma}\sigma^z_{\sigma'\sigma'}]
				\bfS^2_\bfr(t)\cdot\bfS^1_{\bfr'}(t')
\nonumber\\&
			-i[\sigma^z_{\sigma\sigma}-\sigma^z_{\sigma'\sigma'}]
				\bfDelta\cdot[\bfS^2_\bfr(t)\times\bfS^1_{\bfr'}(t')]
		\Bigr\}
	,
\label{eq-dSK}
\end{align}
where the dynamical range functions $F_{\sigma\sigma'}$ are given by
\begin{align}
F_{\sigma\sigma'}(\bfR;\tau)=&
	(-i)\theta(\tau)\int
		[f(\dote{\bfk'\sigma'})-f(\leade{\bfk})]
		e^{i(\bfk-\bfk')\cdot\bfR}
\nonumber\\&\times
		e^{-i(\leade{\bfk}-\dote{\bfk'\sigma'})\tau}
	\frac{d\bfk}{(2\pi)^2}\frac{d\bfk'}{(2\pi)^2}.
\label{eq-Fss'}
\end{align}
Below, we shall calculate those functions explicitly. Before we proceed, however, we note that the electron mediated spin-spin interaction described in Eq. (\ref{eq-dSK}) comprise three different kinds of interactions; Ising, Heisenberg, and Dzyaloshinski-Moriya (DM) types of interactions, respectively, in agreement with Ref. \cite{imamura2004}.

The first term ($\sim[\bfS_\bfr^q(t)\cdot\bfDelta][\bfS_{\bfr'}^c(t')\cdot\bfDelta]$) is a generalized Ising-type of interaction. That it is of Ising-type can be understood since it provides the interaction between the spins projected onto the direction of the field $\bfDelta$. By rotating the reference frame such that e.g. $\bfDelta=\Delta\hat{\bf z}$, one finds that $[\bfS_\bfr^q(t)\cdot\bfDelta][\bfS_{\bfr'}^c(t')\cdot\bfDelta]=\Delta^2S_\bfr^{q,z}(t)S_{\bfr'}^{c,z}(t')$. The Ising interaction vanishes for non-spin polarized conduction electrons, since the range functions are spin-independent, i.e. $F_{\sigma\sigma'}=F$, under such conditions.

The first two contributions are expected to be present between spins interacting via metallic or semi-conducting medium. The last contribution is, on the other hand, expected to arise in anisotropic systems \cite{dzyaloshinski1958,moriya1960}. Here, this anisotropy is generated by the spin polarized surface electrons. For non-chiral spin-polarization of the surface electrons, the range functions $F_{\up\down}=F_{\down\up}$, which leads to that the DM interaction vanishes, as expected.

It is interesting to note, that the DM interaction is non-vanishing whenever the local spins create a spatially inhomogeneous spin-polarization of the surface electrons. In this sense, the induced spin-polarization can be viewed as an effective defect induced spin-orbit interaction.

Next, we calculate the dynamical range functions. The angular integrals in Eq. (\ref{eq-Fss'}) results in the factors $J_0(kR)$, where $J_0(x)$ is the Bessel function. The Fourier transform of the exponential $e^{-i(\leade{\bfk}-\dote{\bfk'\sigma'})\tau}$ is given by $(\omega-\dote{\bfk\sigma}+i\delta)^{-1}(\omega'-\dote{\bfk'\sigma'}-i\delta)^{-1}$. Thus, for quadratic dispersion $\dote{\bfk}=k^2/2N_0$, $N_0=m$, changing momentum to energy integrations, and carrying out those energy integrals, give
\begin{align}
F_{\sigma\sigma'}(\bfR;\tau)=&
	\frac{iN_0^2}{4}
	\theta(\tau)
	e^{-i\Omega_{\sigma\sigma'}\tau}
	\int_{-\infty+i\delta}^{\infty+i\delta}
		[f_{\sigma'}(\omega')-f_\sigma(\omega)]
\nonumber\\&\times
		H_0^{(1)}(\tilde{R}\sqrt{\omega})
		H_0^{(1)}(\tilde{R}\sqrt{\omega'})
		e^{-i(\omega-\omega')\tau}
		\frac{d\omega}{2\pi}\frac{d\omega'}{2\pi},
\end{align}
where $H_n^{(1)}(x)$ is the Hankel function, $\Omega_{\sigma\sigma'}=|\bfDelta|(\sigma-\sigma')/2$, $\tilde{R}=R\sqrt{2N_0}$, and $f_\sigma(\omega)=f(\omega+\sigma|\bfDelta|/2)$. Using that $(-i)\theta(\tau)e^{-ix\tau}=\int(\Omega-x+i\delta)^{-1}e^{-i\Omega\tau}d\Omega/2\pi$, we can finally write $F_{\sigma\sigma'}(\bfR;\tau)=\int F_{\sigma\sigma'}(\bfR,\Omega)e^{-i\Omega\tau}d\Omega/(2\pi)$, where
\begin{align}
F_{\sigma\sigma'}(\bfR,\Omega)=&
	(-i)\frac{N_0^2}{4}
		\int
	[f_\sigma(\omega-\Omega)
	+f_\sigma(\omega)]
	H_0^{(1)}(\tilde{R}\sqrt{\omega})
\nonumber\\&\times
	H_0^{(1)}(\tilde{R}\sqrt{\omega-\Omega+\Omega_{\sigma\sigma'}})
	\frac{d\omega}{2\pi}
\label{eq-Fdyn}
\end{align}
showing that the dynamical range function depends on time $\tau$, the spin bias $\Omega_{\sigma\sigma'}$, and on the spin-chemical potential $\mu_\sigma=\dote{F}-\sigma|\bfDelta|/2$. Eq. (\ref{eq-Fdyn}) is the central result of this paper and below we analyze a few of its consequences to the electron mediated spin-spin exchange interaction.

The static regime, which corresponds to assuming frozen spin moments, is defined for $\Omega=0$. For small spin biases $\Omega_{\sigma\sigma'}/\dote{F}\ll1$, such that $H_0^{(1)}(\tilde{R}\sqrt{\omega+\Omega_{\sigma\sigma'}})\approx H_0^{(1)}(\tilde{R}\sqrt{\omega})$, the integral in Eq. (\ref{eq-Fdyn}) can be analytically calculated for low temperatures ($T\rightarrow0$), for which
\begin{align}
F_{\sigma\sigma'}(\bfR)=&
\frac{N_0}{2\pi}k_\sigma^2\sum_{n=0,1}
	\biggl[
		J_n(Rk_\sigma)Y_n(Rk_\sigma)
\nonumber\\&
		-\frac{i}{2}[
			J_n^2(Rk_\sigma)-Y_n^2(Rk_\sigma)
		]
	\biggr],
\label{eq-Fstat}
\end{align}
where $k_\sigma=\sqrt{k_F^2-\sigma|\bfDelta|N_0}$ with the Fermi vector $k_F=\sqrt{2N_0\dote{F}}$, whereas $Y_n(x)$ is the Neumann function. Here, the real part captures previous results \cite{fischer1975,monod1987,litvinov1998}, in the spin-degenerate limit $k_\sigma\rightarrow k_F$, whereas the imaginary part accounts for the retardation and damping effects of the interaction due to the electron medium.

For $2k_FR\gg1$, the asymptotic expansion of Eq. (\ref{eq-Fstat}) \cite{fischer1975} leads to that $\re F_{\sigma\sigma'}(\bfR)\sim\sin(2k_\sigma R)/(2k_\sigma R)^2$, in agreement with previous results \cite{fischer1975,monod1987,litvinov1998}, which provides the usual spatial decay for the isotropic electron mediated Heisenberg-like exchange, see Eq. (\ref{eq-dSK}). Analogously, the asymptotic expansion of the imaginary part of Eq. (\ref{eq-Fstat}) gives $\im F_{\sigma\sigma'}\sim\cos(2k_ \sigma R)/(2k_ \sigma R)^2$. In contrast, the electron mediated DM interaction asymptotically decays as $\sin(2k_F R)/(2k_F R)$, which can be seen from the following observation. The DM interaction depends on the range functions as $F_{\up\down}-F_{\down\up}$, see Eq. (\ref{eq-dSK}). By Taylor expanding the imaginary part of this difference and using that $k_\sigma\approx k_F-\sigma|\bfDelta|N_0/(2k_F)$, one finds that it asymptotically reduces to
\begin{align}
\im\sum_\sigma\sigma^z_{\sigma\sigma}F_{\up\down}(\bfR)\sim&
	\frac{|\bfDelta|N_0}{2k_F^2}
	\cdot
	\frac{\sin2k_FR}{2k_FR},
\label{eq-DMasympt}
\end{align}
which to the best of our knowledge has not been reported previously. Hence, despite the DM interaction depends quadratically on the anisotropy field $\bfDelta$, it tends to dominate over the Heisenberg, and Ising, exchange for sufficiently large distances between the spins. Thus, in absence of effective magnetic fields acting on the local spins, two spins configure themselves perpendicular to one another when being separated by a sufficiently large distance since the DM interaction dominates their coupling. A collinear alignment of the spins is typically favorable whenever the spins are close to one another, since the Heisenberg interaction provides the strongest contribution to the coupling. The cross over distance at which the DM interaction begins to dominate over the Heisenberg interaction is roughly given by $R_C\approx k_F/(2|\bfDelta|N_0)$. This cross over distance is obtained under the assumption that $N_0|\bfDelta|/k_F^2\ll1$, which is reasonable from the point of view of Ref. \cite{reinert2003}, where $k_F\simeq0.17$ \AA$^{-1}$ whereas the spin-splitting due to the Rashba effect $\sqrt{N_0|\bfDelta|}\simeq0.012$ \AA$^{-1}$ ($|\bfDelta|\sim1$ meV). In terms of these values, the cross over distance $R_C\approx1,180$ \AA. For larger spin-splitting, the above approximations are not valid, however, we expect that the effect of the DM interaction will be even stronger and that the cross over distance is significantly shorter.

\begin{figure}[b]
\begin{center}
\includegraphics[width=0.99\columnwidth]{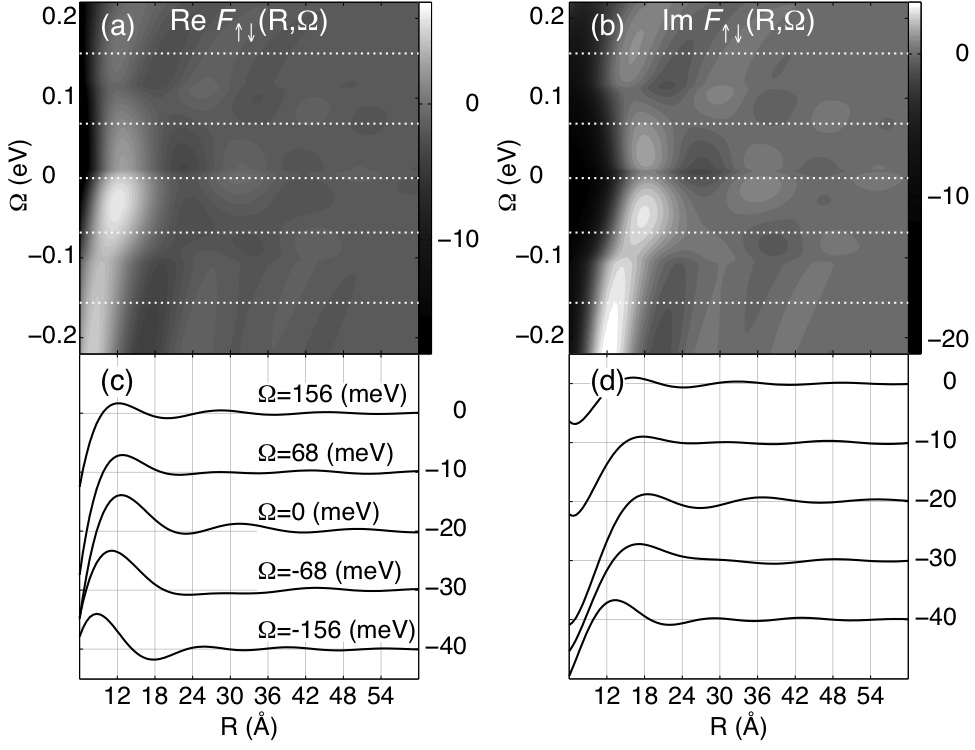}
\end{center}
\caption{Energy ($\Omega$) and spatial ($R$) dependence of the real, (a), (c), and imaginary, (b), (d), parts of $F_{\up\down}(\bfR,\Omega)$. The dotted lines in panels (a) and (b) indicate the energies at which the line scans in panels (c) and (d) are extracted. In panels (c), (d), the plots are vertically shifted for clarity. Here, $k_F=0.17$ \AA$^{-1}$, $\Omega_{\up\down}=10$ meV at $T=10 K$.}
\label{figure1}
\end{figure}

For the remainder of this paper, we consider the local spin moments to be time-dependent, i.e. $\Omega\neq0$, as they would be under the influence of e.g. an effective magnetic field. In Eq. (\ref{eq-dSK}), the dynamical range function $F_{\sigma\sigma'}(\bfR,\tau)$ mediates the interaction between the spins signified by $\bfS^{(1)}(t')$ and $\bfS^{(2)}(t)$, which have different time-arguments. Using that e.g. $\bfS^{(1)}(t)=\int\bfS^{(1)}(x)e^{ixt}dx/(2\pi)$ and integrating over $t$ and $t'$ [c.f. Eq. (\ref{eq-Saction})], we can write, for instance, the Heisenberg type of exchange interaction between two spins located at $\bfr$ and $\bfr'$ as $-(v_uJ_K)^2\sum_{\sigma\sigma'}[1-|\bfDelta|\sigma_{\sigma\sigma}^z\sigma_{\sigma'\sigma'}^z]\int F_{\sigma\sigma'}(\bfR,\Omega)\bfS^{(2)}_\bfr(\Omega)\cdot\bfS^{(1)}_{\bfr'}(-\Omega)d\Omega/(4\pi)$, and analogously for the other two types of interactions. The exchange interaction between the localized spins, hence, strongly depends on the excitation spectra of the spins. In Fig. \ref{figure1}, we plot the real, panels (a), (c), and the imaginary, panels (b), (d), parts of $F_{\sigma\sigma'}(\bfR,\Omega)$ for a given spin bias, $\Omega_{\up\down}\sim10$ meV, and temperature, $T=10$ K. As expected from previous results, and from the above discussion, in the regime near $\Omega=0$, $F_{\sigma\sigma'}$ acquires an oscillatory decaying behavior as function of $R$. 

The expression given in Eq. (\ref{eq-Fdyn}) suggests to interpret the effective exchange interaction as the interference between (spin-dependent) charge density waves with their frequencies set by the spin-chemical potential $\mu_\sigma$, the spin bias $\Omega_{\sigma\sigma'}$, and the excitation spectra of the localized spins. Particularly, the waves are described in terms of the wave vectors $k=\sqrt{2N_0\omega}$ and $k_{\sigma\sigma'}=\sqrt{2N_0(\omega-\Omega+\Omega_{\sigma\sigma'})}$, respectively. Fig. \ref{figure1} (a), (b), display how the oscillatory character of $F_{\sigma\sigma'}(\bfR,\Omega)$, as function of $R$, changes as the energy $\Omega$ varies along the vertical axis. The period of the oscillations is shorter for energies $\Omega\neq0$ compared to the period in the static regime. In the static regime, the frequency of the charge density wave differ only by the spin bias $\Omega_{\sigma\sigma'}$, which typically is small compared to the Fermi energy. Hence, the charge density waves are almost in phase with one another. For finite energies, $\Omega\neq0$, the difference in frequency between the two waves increases, such that the waves go out of phase. Hence, due to the incommensurability of the waves, the resulting period of  $F_{\sigma\sigma'}(\bfR,\Omega)$, as function of $R$, changes for increasing $|\Omega|$.

In conclusion, we have studied the electron mediated spin-spin exchange interaction under non-equilibrium conditions for localized spins embedded in a two-dimensional system. It was demonstrated that the range function depends dynamically on time, the spin excitation spectrum, and the spin-bias between the spin channels in the electron medium. This leads to that the electron mediated exchange interaction between the localized spins is determined by the spin-polarization of the electron medium as well as of the excitation spectra of the local spins. In the case of spatially anisotropic spin-polarized surface electrons, the electron mediated spin-spin exchange interaction comprise the Ising, Heisenberg, and Dzyaloshinski-Moriya type of interactions, capturing the static case previously reported \cite{imamura2004}. It was, moreover, shown that earlier results for the range function, which were derived for systems in the static regime \cite{imamura2004,fischer1975,monod1987,litvinov1998}, can be straightforwardly extended to slowly fluctuating spins. Particularly, the Dzyaloshinski-Moriya interaction was shown to decay as $\sin(2k_FR)/(2k_FR)$ for weakly spin-polarized electrons.

The author thanks A. V. Balatsky, A. Bergman, O. Eriksson, and L. Nordstr\"om for valuable discussions, and for support from the Swedish Research Council.

\end{document}